\documentclass[12pt]{article}%
\usepackage{amsmath}
\usepackage{fancyhdr}
\usepackage{amsfonts}%
\usepackage{amssymb}%
\usepackage{graphicx}

\begin{document}

\title{Quantum Searching amidst Uncertainty}
\author{Lov K. Grover \thanks{Research was partly supported by NSA\ \&\ ARO under
contract DAAG55-98-C-0040.}\\\textit{lkgrover@bell-labs.com}\\Bell Laboratories, Lucent Technologies, \\600-700 Mountain Avenue,\\Murray Hill, NJ 07974}
\date{}
\maketitle

\begin{abstract}
Consider a database most of whose entries are marked but the precise fraction
of marked entries is not known. What is known is that the fraction of marked
entries is $1-\epsilon$, where $\epsilon$ is a random variable that is
uniformly distributed in the range $\left(  0,\epsilon_{0}\right)  .$The
problem is to try to select a marked item from the database in a single query.
If the algorithm selects a marked item, it succeeds, else if it selects an
unmarked item, it makes an error.

How low can we make the probability of error? The best possible classical
algorithm can lower the probability of error to $O\left(  \epsilon_{0}%
^{2}\right)  .$ The best known quantum algorithms for this problem could also
only lower the probability of error to $O\left(  \epsilon_{0}^{2}\right)  .$
Using a recently invented quantum search technique, this paper gives an
algorithm that reduces the probability of error to $O\left(  \epsilon_{0}%
^{3}\right)  .$ The algorithm is asymptotically optimal.

\end{abstract}

\section{Introduction}

Classical search algorithms are robust. If we reduce the problem size, the
algorithm has fewer items to search and the performance of the algorithm will
almost always improve. Quantum search algorithms depend on delicate
interference effects, any change in parameters leads to significantly
different results. For example, if we reduce the number of states in the
database by a factor of four, a quantum search algorithm that would have
previously succeeded, will now fail with certainty. Quantum algorithms usually
exhibit oscillatory behavior in their performance characteristics. As a result
of this, if there is some quantity that we want to maximize, e.g. the
probability of the system being in marked states, we will need very precise
knowledge of the problem parameters. Given such knowledge, it is easy to
fine-tune the algorithm so that it achieves a probability of success of unity.
What happens if we do not have this knowledge? This can be a serious handicap
for a quantum search algorithm. This paper describes a way around this problem.

The original quantum search algorithm \cite{grover96} considered the problem
of finding a marked item in a large unsorted database with minimum queries to
the database. For this type of problem it is usually enough to be able to
obtain the correct answer with a constant probability since the procedure can
be repeated a logarithmic number of times to drive the probability
exponentially close to unity. In that problem a logarithmic factor was not
significant since the quantum search algorithm gave a square-root asymptotic
improvement. However there are other important problems where the additional
queries create a significant overhead and need to be minimized, e.g. when we
are limited to a single query and have to find the answer with a probability
approaching unity. One field in which this type of problem occurs is in
pattern recognition and image analysis where each query requires a lot of
signal processing and the consequences of making an error are catastrophic.

\section{The problem}

Consider the situation where a large fraction of the items in a database are
marked, but the precise fraction of marked items is not known. The goal is to
find a single marked item with as high a probability as possible in a single
query to the database. For concreteness, say some unknown fraction
$(1-\epsilon)$ of the items are marked, with $\epsilon$ uniformly distributed
\ in the range $\left(  0,\epsilon_{0}\right)  $ with equal probability. The
search algorithm returns an item, if it is a marked item the algorithm is said
to succeed, else if it is unmarked, the algorithm is said to fail.

Classically the smallest error that can be obtained is $O\left(  \epsilon
_{0}^{2}\right)  $. To achieve this, uniformly guess an input and evaluate the
function. If the function outputs 1, you're done, otherwise guess again and
cross your fingers (because you have no more evaluations left). One cannot do
better than $O\left(  \epsilon_{0}^{2}\right)  $ classically. In this paper we
show that the probability of failure for the new scheme is $O\left(
\epsilon_{0}^{3}\right)  ,$ whereas that of the best (possible) classical
scheme and that of the best known quantum schemes are both $O\left(
\epsilon_{0}^{2}\right)  $.

Before considering the specific problem mentioned above, let us describe a
general framework. Consider the following transformation%
\begin{align}
& UR_{s}U^{\dagger}R_{t}U\left\vert s\right\rangle \label{transf}\\
R_{s}  & =I-(1-\exp\left(  i\frac{\pi}{3}\right)  )\left\vert s\right\rangle
\left\langle s\right\vert ,\ \ \ R_{t}=I-(1-\exp\left(  i\frac{\pi}{3}\right)
)\left\vert t\right\rangle \left\langle t\right\vert \nonumber
\end{align}
$U$ is an arbitrary unitary transformation, $R_{t}$ \& $R_{s}$ denote
selective phase shifts of the respective state(s) by $\frac{\pi}{3}.$ Note
that if we were to change these phase shifts from $\frac{\pi}{3}$ to $\pi
,$\ we would get one iteration of the amplitude amplification algorithm
\cite{bht}, \cite{ampt. amp.}.

The next section shows that if $U$ drives the state vector from a source
$\left(  s\right)  $ to a target $\left(  t\right)  $ state with a probability
of $\left(  1-\epsilon\right)  $, i.e. $\left\Vert U_{ts}\right\Vert
^{2}=\left(  1-\epsilon\right)  $, then the transformation (\ref{transf})
drives the state vector from the source to the same target state with a
probability of $\left(  1-\epsilon^{3}\right)  .$ The deviation from the $t$
state has hence fallen from $\epsilon$\ to $\epsilon^{3}$. Note that this is
different from the amplitude amplification framework where the
\textit{amplitudes} were getting amplified; over here it is more convenient to
present the results in terms of the \textit{probabilities}.

The striking aspect of this result is that it holds for \textit{any} kind of
deviation from the $t$\ state. Unlike the standard search (or amplitude
amplification) algorithm which would greatly overshoot the target state when
$\epsilon$ is small (Figure 1); the new algorithm will always move towards the
target. As shown in Section 5, this feature of the framework can be used to
develop algorithms that are more robust to variations in the problem parameters.%

\begin{figure}
[ptb]
\begin{center}
\includegraphics[
trim=0.000000in 1.154188in 0.000000in 3.350821in,
height=1.5887in,
width=5.2399in
]%
{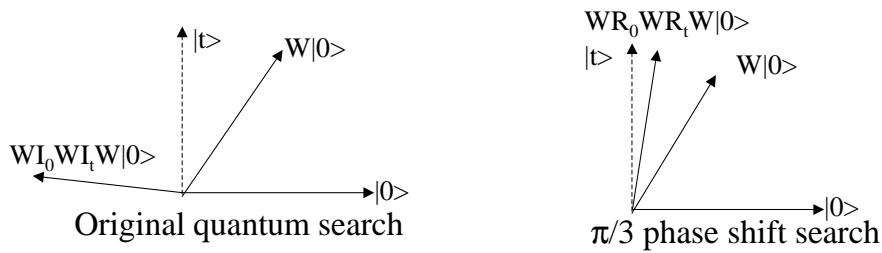}%
\caption{The original quantum search algorithm is very sensitive to the
fraction of marked states. For example, when the fraction is about 3/4, the
algorithm fails, the algorithm of this paper will always produce an
improvement.}%
\end{center}
\end{figure}

\section{Analysis}

We analyze the effect of the transformation $UR_{s}U^{\dagger}R_{t}U$ when it
is applied to the $\left\vert s\right\rangle $ state. As mentioned in the
previous section, $R_{t}$ \& $R_{s}$ denote selective phase shifts of the
respective state(s) by $\frac{\pi}{3}$ ($t$ for target, $s$ for source). We
show that if $\left\Vert U_{ts}\right\Vert ^{2}=\left(  1-\epsilon\right)  ,$
then
\begin{equation}
\left\Vert \left\langle t\right\vert UR_{s}U^{\dagger}R_{t}U\left\vert
s\right\rangle \right\Vert ^{2}=\left(  1-\epsilon^{3}\right)  .
\end{equation}

In the rest of this section, the greek alphabet $\theta$ will be used to
denote $\frac{\pi}{3}.$ Start with $\left\vert s\right\rangle $ and apply the
operations $U,R_{s},U^{\dagger},R_{t}$ \&$\ U.$ If we analyze the effect of
the operations, one by one, just as in the original quantum search algorithm
\cite{grover96}, we find that it leads to the following superposition (this
calculation is carried out in the appendix):%
\[
U\left\vert s\right\rangle \left(  e^{i\theta}+\left\Vert U_{ts}\right\Vert
^{2}\left(  e^{i\theta}-1\right)  ^{2}\right)  +\left\vert t\right\rangle
U_{ts}\left(  e^{i\theta}-1\right)  .
\]

To calculate the deviation of this superposition from $\left\vert
t\right\rangle $, consider the amplitude of the above superposition in
non-target states. The probability is given by the absolute square of the
corresponding amplitude:%
\[
\left(  1-\left\Vert U_{ts}\right\Vert ^{2}\right)  \left\Vert \left(
e^{i\theta}+\left\Vert U_{ts}\right\Vert ^{2}\left(  e^{i\theta}-1\right)
^{2}\right)  \right\Vert ^{2}.
\]

Substituting $\left\Vert U_{ts}\right\Vert ^{2}=\left(  1-\epsilon\right)  ,$
the above quantity becomes$:$%
\begin{align*}
&  \epsilon\left\Vert \left(  e^{i\theta}+\left(  1-\epsilon\right)  \left(
e^{i\theta}-1\right)  ^{2}\right)  \right\Vert ^{2}\\
&  =\epsilon\left\Vert \left(  -e^{i\theta}+e^{2i\theta}+1\right)
-\epsilon\left(  e^{i\theta}-1\right)  ^{2}\right\Vert ^{2}\\
&  =\epsilon^{3}.
\end{align*}

\section{Existing Algorithms}

\subsection{Classical Algorithm}

The best classical algorithm is to select a random state and see if it is a
$t$\ state (one query). If yes, return this state; if not, pick another random
state and return that without any querying. Note that this algorithm requires
a single query, not two. The probability of failure is equal to that of not
getting a single $t$\ state in two random picks since if either of the two
states is a marked state, the algorithm will succeed. This probability is
equal to $\epsilon^{2}.$ Since $\epsilon$ is uniformly distributed in the
range $\left(  0,\epsilon_{0}\right)  $. The overall failure probability
becomes$\frac{\int_{0}^{\epsilon_{0}}\epsilon^{2}d\epsilon}{\int_{0}%
^{\epsilon_{0}}d\epsilon}=\frac{1}{3}\epsilon_{0}^{2}$
\begin{figure}
[ptb]
\begin{center}
\includegraphics[
trim=0.974980in 2.744853in 0.000000in 1.863651in,
height=1.5091in,
width=4.6484in
]%
{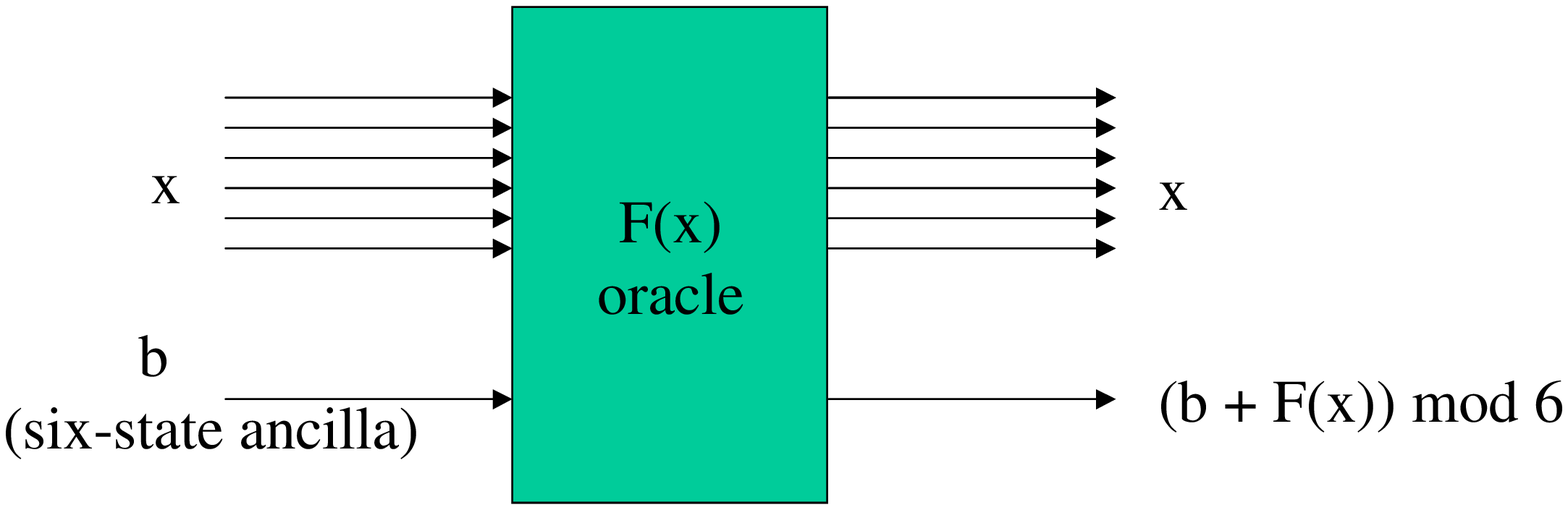}%
\caption{By setting the six-state ancilla, b, to the superposition $\frac
{1}{\sqrt{6}}\left(  \left\vert 0\right\rangle +\left\vert 1\right\rangle
\omega+\left\vert 2\right\rangle \omega^{2}+\left\vert 3\right\rangle
\omega^{3}+\left\vert 4\right\rangle \omega^{4}+\left\vert 5\right\rangle
\omega^{5}\right)  $ where $\omega=\exp\left(  -\frac{i\pi}{3}\right)  ,$ we
get a $\frac{\pi}{3}$ phase-shift of the states for which $F(x)=1$ relative to
those for which $F\left(  x\right)  =0.$ $\cite{tath}$ gives a technique for
implementing this with qubits.}%
\end{center}
\end{figure}

\subsection{Quantum Searching}

Boyer, Brassard, H\o yer and Tapp \cite{bbht} first described in detail an
algorithm that succeeds with probability approaching 1, regardless of the
number of solutions (it's a classical algorithm that uses quantum searching as
a subroutine; of course, it can be made fully quantum.) The first quantum
algorithm to be able to search in a single query with a success probability
approaching 1 was given by Mosca \cite{mosca}.

Mosca observed that the quantum counting algorithm of \cite{bbht} (based on
the original searching algorithm) produces a solution with probability
converging to $1/2$. One easily converts this to an algorithm with probability
of success converging to $1/4$. Thus by using this algorithm as a sub-routine
in another quantum search, we get success probability converging to $1$. (This
appears to be based on the observation in \cite{bbht} that an algorithm that
succeeds with probability exactly $1/4$ can be amplified to one with success
probability exactly $1$ using only one quantum search iterate). In other
words, the technique Mosca uses is to take a search algorithm that succeeds
with probability $1/4-X$ and then use one quantum search iteration to map it
to an algorithm that succeeds with probability $\left(  1-12X^{2}%
-16X^{3}\right)  $. Using this scheme, if the fraction of marked states of the
database is $1-\epsilon$, one can easily obtain a marked state with a
probability of error of $1-\frac{3}{4}\epsilon^{2}-\frac{1}{4}\epsilon^{3}$ by
means of a single quantum query. The overall failure probability in this case
becomes $\frac{\int_{0}^{\epsilon_{0}}\left(  \frac{3}{4}\epsilon^{2}+\frac
{1}{4}\epsilon^{3}\right)  d\epsilon}{\int_{0}^{\epsilon_{0}}d\epsilon}%
=\frac{1}{4}\epsilon_{0}^{2}+\frac{1}{16}\epsilon_{0}^{3}$

A recent quantum search based algorithm for this problem is by Younes et al
\cite{younes} (actually it is somewhat unfair to compare it to the other
algorithms since this was specifically designed to perform well when the
success probability was close to $\frac{1}{2},$ not close to $1)$. This finds
a solution with a probability of $\left(  1-\cos\theta\right)  \left(
\frac{\sin^{2}\left(  q+1\right)  \theta}{\sin^{2}\theta}+\frac{\sin
^{2}q\theta}{\sin^{2}\theta}\right)  ,$ where $q$ = number of queries and
$\theta=\arccos\epsilon$ ((59) from \cite{younes})$.$ When $q=1,$ the success
probability becomes: $\left(  1-\epsilon\right)  \left(  1+4\epsilon
^{2}\right)  ,$ hence the probability of error becomes $\epsilon-4\epsilon
^{2}+4\epsilon^{3}$. The overall failure probability becomes $\frac{\int
_{0}^{\epsilon_{0}}\left(  \epsilon-4\epsilon^{2}+4\epsilon^{3}\right)
d\epsilon}{\int_{0}^{\epsilon_{0}}d\epsilon}=\left(  \frac{1}{2}\epsilon
_{0}-\frac{4}{3}\epsilon_{0}^{2}+\epsilon_{0}^{3}\right)  $

\section{New algorithm}

As in the quantum search algorithm, encode the $N$ items in the database in
terms of $\log_{2}N$ qubits. The algorithm consists of applying the
transformation $WR_{\overline{0}}WR_{t}W$ to $\left\vert \overline
{0}\right\rangle $. W is the Walsh-Hadamard Transformation and $\overline{0}$
is the state with all qubits in the $0$ state. After this an observation is
made which makes the system collapse into a basis state.

In order to analyze the performance of this algorithm, note that the algorithm
is merely the phase shift transformation $UR_{s}U^{\dagger}R_{t}U$ applied to
\ $\left\vert s\right\rangle $ which has already been analyzed in section 3.
$U$\ is the W-H transform ($W$) and the state $s$\ is the $\overline{0}$ state
(state with all qubits in the 0 state), then $\left\Vert U_{ts}\right\Vert
^{2}=1-\epsilon,$ where $\epsilon$ lies in the range $\left(  0,\epsilon
_{0}\right)  .$ Therefore after applying the transformation $WR_{\overline{0}%
}WR_{t}W$ to $\left\vert \overline{0}\right\rangle $\ , the probability of
being in a non-$t$\ state becomes $\epsilon^{3}$, i.e. the overall failure
probability becomes $\frac{\int_{0}^{\epsilon_{0}}\epsilon^{3}d\epsilon}%
{\int_{0}^{\epsilon_{0}}d\epsilon}=\frac{1}{4}\epsilon_{0}^{3}$
\begin{figure}
[ptb]
\begin{center}
\includegraphics[
trim=0.000000in 1.617664in 0.000000in 1.848652in,
height=2.13in,
width=5.2399in
]%
{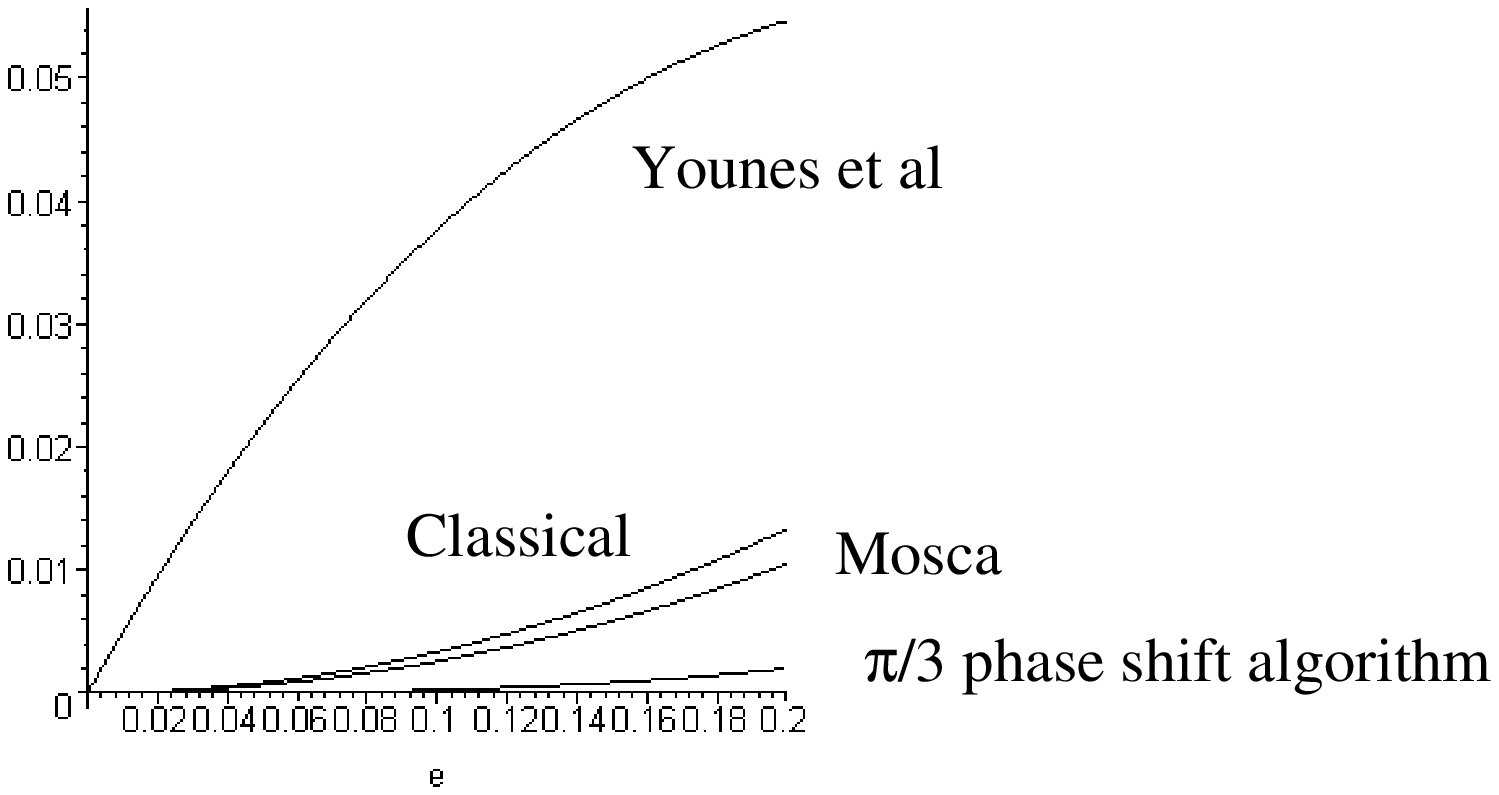}%
\caption{Comparison of the failure probability of the $\pi/3$ phase shift
algorithm with a classical algorithm, \cite{younes}, and \cite{mosca}, when
the fraction of unmarked states $\epsilon$, varies between 0 \& 0.2.}%
\end{center}
\end{figure}

The performance of the algorithm is graphically illustrated in Figure 3.

\section{Extensions}

\subsection{Multi-query searching}

In practice, a database search would use multiple queries. The technique
discussed above extends neatly to the multi-query situation. As described in
\cite{quantph}, the multi-query algorithm based on the $\frac{\pi}{3}$
phase-shift transformation is able to reduce the probability of error to
$\frac{\epsilon_{0}^{2q+1}}{2q+2}$ after $q$ queries to the database.
A\ classical algorithm reduces the probability of error to $\frac{\epsilon
_{0}^{q+1}}{q+2}.$ Note that $q=1$ gives the same results as in section 5.

\subsection{Optimality}

Both the single query, as well as the multi-query algorithms are
asymptotically optimal in the limit of small $\epsilon_{0}$. This is
separately proved in \cite{pi_3_srch}.

\subsection{Quantum Control \& Error Correction}

Connections to control and error correction might be evident. Let us say that
we are trying to drive a system from an $s$\ to a $t$\ state/subspace. The
transformation that we have available for this is $U$ which drives it from
$s$\ to $t$\ with a probability $\left\Vert U_{ts}\right\Vert ^{2}$ of
$\left(  1-\epsilon\right)  ,$ i.e. the probability of error in this
transformation is $\epsilon$. Then the composite transformation $UR_{s}%
U^{\dagger}R_{t}U$\ will reduce the error to $\epsilon^{3}.$

This technique is applicable whenever the transformations $U,\ U^{\dagger
},R_{s}\ \&\ R_{t}$ can be implemented. This will be the case when errors are
systematic errors or slowly varying errors, e.g. due to environmental
degradation of some component. This would not apply to errors that come about
as a result of sudden disturbances from the environment. It is further assumed
that the transformation $U$ can be inverted with exactly the same error.
Traditionally quantum error correction is carried out at the single qubit
level where individual errors are corrected, each error being corrected in a
separate way. With the machinery of this paper, errors can be corrected
without ever needing to identify the error syndrome. This is discussed in
\cite{quantph} and \cite{reich}.

\section{Appendix}

\bigskip We analyze the effect of the transformation (3) of section 3 when it
is applied to the $\left\vert s\right\rangle $ state. As in section 3,
$\theta$ will denote $\frac{\pi}{3}.$ We show that if $\left\Vert
U_{ts}\right\Vert ^{2}=\left(  1-\epsilon\right)  ,$ then
\begin{equation}
\left\Vert \left\langle t\right\vert UR_{s}U^{^{^{\dag}}}R_{t}U\left\vert
s\right\rangle \right\Vert ^{2}=\left(  1-\epsilon^{3}\right)
\end{equation}

\begin{description}
\item[starting state] $\left\vert s\right\rangle \qquad\qquad\qquad
\qquad\qquad\qquad\qquad\qquad\qquad\qquad\qquad\qquad\qquad\qquad\qquad
\qquad\qquad\qquad\qquad\qquad\qquad\qquad\qquad\qquad\qquad\qquad\qquad
\qquad\qquad\qquad\qquad\qquad\qquad\qquad\qquad\qquad\qquad\qquad\qquad
\qquad\qquad\qquad\qquad\qquad$

\item[after $R_{t}U$]
\[
R_{t}U\left\vert s\right\rangle =U_{ts}\left(  e^{i\theta}-1\right)
\left\vert t\right\rangle +U\left\vert s\right\rangle
\]

\item[after $U^{^{^{\dag}}}:$]
\[
U^{^{^{\dag}}}R_{t}U\left\vert s\right\rangle =\left\vert s\right\rangle
+U_{ts}\left(  e^{i\theta}-1\right)  U^{^{^{\dag}}}\left\vert t\right\rangle
\]

\item[after $R_{s}$] $:$%
\begin{align*}
R_{s}U^{^{^{\dag}}}R_{t}U\left\vert s\right\rangle  & =\left(  \left\vert
s\right\rangle +U_{ts}\left(  e^{i\theta}-1\right)  U^{^{^{\dag}}}\left\vert
t\right\rangle \right) \\
& +\left(  e^{i\theta}-1\right)  \left(  \left\vert s\right\rangle
+U_{ts}\left(  e^{i\theta}-1\right)  \left\vert s\right\rangle \left\langle
s\right\vert U^{^{^{\dag}}}\left\vert t\right\rangle \right) \\
& =\left\vert s\right\rangle \left(  e^{i\theta}+\left\Vert U_{ts}\right\Vert
^{2}\left(  e^{i\theta}-1\right)  ^{2}\right)  +U_{ts}\left(  e^{i\theta
}-1\right)  U^{^{^{\dag}}}\left\vert t\right\rangle
\end{align*}

\item[after the final $U$:]
\[
UR_{s}U^{^{^{\dag}}}R_{t}U\left\vert s\right\rangle =U\left\vert
s\right\rangle \left(  e^{i\theta}+\left\Vert U_{ts}\right\Vert ^{2}\left(
e^{i\theta}-1\right)  ^{2}\right)  +U_{ts}\left(  e^{i\theta}-1\right)
\left\vert t\right\rangle
\]

\item[Calculating the error] To estimate the error, consider the amplitude of
the above superposition in non-target states (this is due to the portion
$U\left\vert s\right\rangle \left(  e^{i\theta}+\left\Vert U_{ts}\right\Vert
^{2}\left(  e^{i\theta}-1\right)  ^{2}\right)  ,$ the other portion
$U_{ts}\left(  e^{i\theta}-1\right)  \left\vert t\right\rangle $ is clearly in
the target state). The magnitude of the error probability is given by the
absolute square of this error amplitude:%
\[
\left(  1-\left\Vert U_{ts}\right\Vert ^{2}\right)  \left\Vert \left(
e^{i\theta}+\left\Vert U_{ts}\right\Vert ^{2}\left(  e^{i\theta}-1\right)
^{2}\right)  \right\Vert ^{2}%
\]

\end{description}

Assume $\left\Vert U_{ts}\right\Vert ^{2}$ to be $\left(  1-\epsilon\right)  ,
$ the above quantity becomes$:$%
\begin{align*}
& \epsilon\left\Vert \left(  e^{i\theta}+\left(  1-\epsilon\right)  \left(
e^{i\theta}-1\right)  ^{2}\right)  \right\Vert ^{2}\\
& =\epsilon\left\Vert \left(  -e^{i\theta}+e^{2i\theta}+1\right)
-\epsilon\left(  e^{i\theta}-1\right)  ^{2}\right\Vert ^{2}\\
& =\epsilon^{3}%
\end{align*}

\end{document}